\documentclass[a4paper,11pt]{article}

\usepackage{amsmath}
\usepackage{amsfonts}
\usepackage{amssymb}
\usepackage{caption}
\usepackage{graphicx}
\usepackage{graphics}
\usepackage{color}
\usepackage{ushort}
\usepackage{makeidx}
\usepackage{float}
\usepackage{appendix}
\usepackage{accents}
\usepackage{framed}
\usepackage{versions}
\usepackage{xcolor}
\usepackage{mathbbol}
\usepackage{blindtext}
\usepackage{emptypage}
\usepackage{verbatim}
\usepackage{enumitem}
\usepackage{array}
\usepackage{multirow}

\def\hmath$#1${\texorpdfstring{{\rmfamily\textit{#1}}}{#1}}

\usepackage[bookmarks]{hyperref}
\hypersetup{
    colorlinks=true,   	
    linkcolor=red,      
    citecolor = [rgb]{0 0.7 0},   	
    filecolor=magenta, 	
    urlcolor=blue
}

\newcommand{\IND}{{\mathbb I}}

\def\ba{\begin{align}}
\def\ea{\end{align}}
\def\ban{\begin{align*}}
\def\ean{\end{align*}}

\def\be{\begin{eqnarray}}
\def\ee{\end{eqnarray}}
\def\ben{\begin{eqnarray*}}
\def\een{\end{eqnarray*}}

\def\bqq{\begin{equation}}
\def\eqq{\end{equation}}
\def\bqqn{\begin{equation*}}
\def\eqqn{\end{equation*}}

\def\elabel#1{\label{e:#1}}

\def\sq{$\Box$}

\def\qed{\ifmmode\sq\else{\unskip\nobreak\hfil
\penalty50\hskip1em\null\nobreak\hfil\sq
\parfillskip=0pt\finalhyphendemerits=0\endgraf}\fi\par\medbreak}

\newsavebox{\junk}
\savebox{\junk}[1.6mm]{\hbox{$|\!|\!|$}}

\def\til={{\widetilde =}}

 \def\eq#1/{(\ref{#1})}

\def\eq#1/{(\ref{e:#1})}

\newcommand{\beqn}[1]{\notes{#1}%
\begin{eqnarray} \elabel{#1}}

\newcommand{\eeqn}{\end{eqnarray} } 

\newcommand{\beq}[1]{\notes{#1}%
\begin{equation}\elabel{#1}}

\newcommand{\eeq}{\end{equation}} 

\def\bdes{\begin{description}}
\def\edes{\end{description}}

\def\notes#1{}

\definecolor{mag}{rgb}{0.7,0,0.3}
\definecolor{dgreen}{rgb}{0.1,0.5,0.1}
\definecolor{dred}{rgb}{.8,0,0}
\definecolor{gray}{rgb}{.8,.8,.8}
\definecolor{brown}{rgb}{0.6451,0.3706,0.1745}

\renewcommand{\ref}[1]{(\ref{#1})}

\newcommand{\PP} { \mathbb{P} }
\newcommand{\EE} { \mathbb{E} }

\begin{document}
 
\title{\vspace{-1.1cm}%
An Information-Theoretic Proof of a Finite de Finetti Theorem}

\author
{
	Lampros Gavalakis
    \thanks{Department of Engineering,
	University of Cambridge,
        Trumpington Street,
	Cambridge CB2 1PZ, U.K.
                Email: \texttt{\href{mailto:lg560@cam.ac.uk}%
			{lg560@cam.ac.uk}}.
	L.G.\ was supported in part by EPSRC grant number RG94782.
        }
\and
        Ioannis Kontoyiannis 
    \thanks{Statistical Laboratory, DPMMS,
	University of Cambridge,
	Centre for Mathematical Sciences,
        Wilberforce Road,
	Cambridge CB3 0WB, U.K.
                Email: \texttt{\href{mailto:yiannis@maths.cam.ac.uk}%
			{yiannis@maths.cam.ac.uk}}.
		Web: \texttt{\url{http://www.dpmms.cam.ac.uk/person/ik355}}.
	I.K.\ was supported in part by the Hellenic Foundation for Research 
	and Innovation (H.F.R.I.) under the ``First Call for H.F.R.I. Research 
	Projects to support Faculty members and Researchers and the 
	procurement of high-cost research equipment grant,'' project 
	number 1034.
        }
}

\date{\today}

\maketitle

\begin{abstract}
A finite form of de Finetti's representation theorem
is established using elementary information-theoretic tools: 
The distribution of the first $k$ random variables in an
exchangeable binary vector of length $n\geq k$
is close to a mixture of product distributions. 
Closeness is measured in terms of the relative entropy 
and an explicit bound is provided.
\end{abstract}


\section{Introduction}

A finite sequence of random variables $(X_1,X_2,\ldots,X_n)$ is 
{\em exchangeable} if it has the same distribution
as $(X_{\pi(1)},X_{\pi(2)},\ldots,X_{\pi(n)})$ 
for every permutation $\pi$ of $\{1,2,\ldots,n\}$. 
An infinite sequence $\{X_k\;;k\geq 1\}$ is 
exchangeable if  $(X_1,X_2,\ldots,X_n)$ is
exchangeable for all $n$. The celebrated representation
theorem of de Finetti 
\cite{definetti:31,definetti:37}
states that the distribution
of any infinite exchangeable sequence 
of binary random variables can be expressed
as a mixture of the distributions
corresponding to independent and identically distributed
(i.i.d.) Bernoulli trials. For discussions of the 
role of de Finetti's theorem in connection with
the foundations of Bayesian statistics and
subjective probability see, e.g,
\cite{diaconis:77,bayarri:04}
and the references therein.
 
Although it is easy to see via simple examples that
de Finetti's theorem may fail for finite binary 
exchangeable sequences, for large
but finite $n$ the distribution of the first $k$ 
random variables of an exchangeable vector of length
$n$ admits an approximate de~Finetti-style representation.
Quantitative versions of this statement have been
established by Diaconis~\cite{diaconis:77}
and Diaconis and Freedman~\cite{diaconis-freedman:80b}.
The approach of Diaconis' proof in~\cite{diaconis:77}
is based on a geometric interpretation of the set of exchangeable 
measures as a convex subset of the probability simplex.

The purpose of this note is to provide a new
information-theoretic proof of a related finite version
of de Finetti's theorem. For each $p\in[0,1]$,
let $P_p$ denote the Bernoulli probability mass
function with parameter $p$, $P_p(1)=1-P_p(0)=p$,
and write $D(P\|Q)=\sum_{x\in B}P(x)\log[P(x)/Q(x)]$
for the relative entropy (or Kullback-Leibler divergence)
between two probability mass functions $P,Q$ on the same
discrete set $B$;
throughout, 
`log' denotes the natural
logarithm to base $e$.

\medskip
\noindent
{\bf Theorem.}
{\em Let $n \geq 2$. If the binary random variables
$(X_1,X_2,\ldots,X_n)$ are exchangeable, 
then there is a probability measure 
$\mu$ on $[0,1]$, such that, for every $1\leq k \leq n$,
the relative entropy
between the probability mass function $Q_k$ of $(X_1,X_2,\ldots,X_k)$
and the mixture $M_{k,\mu}:=\int P_p^kd\mu(p)$
satisfies:}
\begin{equation}
\label{eq:Dbound}
D(Q_{k}\|M_{k,\mu}) \leq \frac{5k^2\log n}{n-k}.
\end{equation}

By Pinsker's inequality 
\cite{csiszar:67,kullback:67},
$\|P-Q\|^2\leq 2D(P\|Q)$,
the theorem also implies that,
\begin{equation} 
\label{eq:l1bound}
\|Q_k - M_{k,\mu} \| 
\leq k\Bigl(\frac{10\log{n}}{n-k}\Bigr)^{\frac{1}{2}},
\end{equation}
where $\|P-Q\|:=2\sup_B |P(B)-Q(B)|$ 
denotes the total variation distance
between $P$ and~$Q$.
This bound is suboptimal in that,
as shown by Diaconis and Freedman~\cite{diaconis-freedman:80b},
the correct rate with respect to the total variation 
distance in~(\ref{eq:l1bound}) is $O(k/n)$.
On the other hand,~(\ref{eq:Dbound}) gives
an explicit bound for the stronger notion
of relative entropy `distance.'

Rather than to obtain optimal rates,
our primary motivation is 
to illustrate how elementary
information-theoretic ideas can be used
to provide an alternative proof strategy for
de Finetti's theorem, following
a long series of works developing this point of view,
including information-theoretic proofs of
Markov chain convergence~\cite{renyi:61,kendall:63},
the central limit theorem~\cite{barron:clt,artstein:04},
Poisson and compound Poisson 
approximation~\cite{konto-H-J:05,Kcompound:10},
and the Hewitt-Savage 0-1 law~\cite{oconnell:TR}.

Before turning to the proof, we mention that
there are numerous generalisations and extensions of
de Finetti's classical theorem and its finite version
along different directions; see, 
e.g.,~\cite{diaconis:88}
and the references therein.
The classical de Finetti representation theorem
has been shown to hold for exchangeable processes
with values in much more general spaces than 
$\{0,1\}$~\cite{hewitt-savage:55}, and 
for mixtures of Markov chains~\cite{diaconis-freedman:80}.
Recently, an elementary proof of de Finetti's theorem 
for the binary case was given in~\cite{kirsch:19},
a more analytic proof appeared in~\cite{alam:20},
and connections with category theory were drawn
in~\cite{fritz}.

\section{Proof of the finite de Finetti theorem}

We first need to introduce some notation.
Let $n\geq 2$ be fixed.
For any $1\leq i\leq j\leq n$, write $X_i^j$
for the block of random variables $X_i^j=(X_i,X_{i+1},\ldots,X_j)$.
Denote by $N_{i,j}$ the number of 1s in $X_i^j$,
so that $N_{i,j}=\sum_{k=i}^jX_k$, and for every $0\leq\ell\leq n$
write $A_\ell$ for the event $\{N_{1,n}=\ell\}$.

The main step of the proof is the estimate in the lemma below,
which gives a bound on the degree of dependence
between $X_i$ and $X_{i+1}^k$, conditional on $A_\ell$.
This bound is expressed in terms of the mutual information.
Let $(X,Y)$ be two discrete random variables with joint
probability mass function (p.m.f.) $P_{XY}$ and marginal
p.m.f.s $P_X$ and $P_Y$, respectively. Recall that
the entropy $H(X)$ of $X$, often viewed as a measure
of the inherent ``randomness'' of $X$~\cite{cover:book2},
is defined as,
$H(X)=H(P_X)=-\sum_{x}P_X(x)\log P_X(x)$, 
where the sum is over all possible values of $X$
with nonzero probability. Similarly, the conditional
entropy of $Y$ given $X$ is,
$$H(Y|X)=-\sum_x P_X(x)\sum_y P_{Y|X}(y|x)\log 
P_{Y|X}(y|x),$$
where $P_{Y|X}(y|x)=P_{XY}(x,y)/P_X(x)$.

The mutual information between $X$ and $Y$
is $I(X;Y) = H(Y) - H(Y|X)$, and it can also be expressed
as:
$$I(X;Y) = H(X) - H(X|Y)=H(X)+H(Y)-H(X,Y)=D(P_{XY}\|P_XP_Y).$$
For any event $A$, we write $I(X;Y|A)$ for the
mutual information between $X$ and $Y$ when all
relevant p.m.f.s are conditioned on $A$.

From the definition, 
an obvious interpretation of $I(X;Y)$ 
is as a measure of the amount of ``common randomness''
in $X$ and $Y$. Additionally, since $I(X;Y)$
is always nonnegative and equal to zero iff $X$ and $Y$
are independent, the mutual information can be viewed
as a universal, nonlinear measure of dependence between
$X$ and $Y$.
See \cite{cover:book2} for standard properties of
the entropy, relative entropy and mutual information.

Finally, we record an elementary bound that will
be used in the proof of the lemma.
Write $h(p)=-p\log p-(1-p)\log(1-p)$,
$p\in[0,1]$, for the binary entropy function.
Then a simple Taylor expansion gives:
\be
| h (p) - h (q)| \leq |p-q| \times\max\left\{
\Big| \log \Big(\frac{1-p}{p}\Big) \Big|,
\Big| \log \Big(\frac{1-q}{q}\Big) \Big|\right\},
\qquad p,q\in(0,1).
\label{eq:SV}
\ee

\medskip

\noindent
{\bf Lemma.}
For all $1\leq k\leq n$, all  $1 \leq i \leq k-1$, and any 
$0\leq\ell\leq n$:
$$
I(X_i;X_{i+1}^k|A_\ell) \leq \frac{5k\log n}{n-k}.
$$

\noindent
{\sc Proof.}
We assume without loss of generality that $k\leq n/2$,
for otherwise the result is trivially true since the
mutual information in the statement is always no greater than~1.
Also, if $\ell=0$ or~$n$ the conditional mutual information
is zero and the result is again trivially true.
Let $Q_n$ denote the p.m.f.\ of $X_1^n$.
By exchangeability, conditional on 
$A_\ell$, all sequences in $\{0,1\}^n$ with 
exactly $\ell$ 1s have the same probability 
under $Q_n$, so $X_1^n$ conditional on $A_\ell$
is uniformly distributed among all such sequences. 
This implies that for all $1 \leq k \leq n/2$, $1\leq i \leq k-1,$
and $1\leq\ell\leq n-1$,
\begin{equation*} 
\PP(X_i = 1| N_{1,n} = \ell, N_{i+1,k})
= \frac{\binom{n-(k-i)-1}{\ell- N_{i+1,k}-1}}{\binom{n-(k-i)}{\ell-N_{i+1,k}}}
= \frac{\ell-N_{i+1,k}}{n-(k-i)}.
\end{equation*}
For the mutual information we have:
\be 
I(X_i;X_{i+1}^k|A_\ell)
&=& \EE\left(\left.h\Big(\frac{\ell}{n}\Big) 
	- h\Big(\frac{\ell-N_{i+1,k}}{n - (k-i)}\Big)\right|
	N_{1,n} = \ell\right)
	\nonumber\\
&\leq&
	 \EE\left(\left.
	\Big|h\Big(\frac{\ell}{n}\Big) 
	- h\Big(\frac{\ell-N_{i+1,k}}{n- (k-i)}\Big)\Big|
	\IND_{\{\ell + k - i - n + 1 
	\leq N_{i+1,k} < \ell\}}
	\right|N_{1,n} = \ell\right) 
	\nonumber \\ 
&+& 
	h\Big(\frac{\ell}{n}\Big)
	\PP(N_{i+1,k} = \ell \big|A_\ell)
	 + h\Bigl(\frac{\ell}{n}\Big)
	\PP(N_{i+1,k} \leq \ell+k-i-n \big| A_\ell).
	\label{threeterms}
\ee
If the probability in the third term above is nonzero, 
then necessarily 
$\ell \geq n - k + 1$ and thus, using 
$n \geq 2k$, $h(\ell/n) \leq 2\frac{k}{n-k}\log{n}$. 
On the other hand, if $\ell + k - i - n + 1 \leq N_{i+1,k} < \ell$,
then 
both $\ell/n$ and $(\ell-N_{i+1,k})/(n-(k-i))$ are between
$1/n$ and $(n-1)/n$, so 
from~(\ref{eq:SV})
the first term 
in \eqref{threeterms} is bounded above by,
$$
\frac{\ell(k-i) + n\EE(N_{i+1,k}|N_{1,n} = \ell)}{n(n-(k-i))}\log{n} 
= \frac{2\ell(k-i)}{n(n-(k-i))}\log{n}.
$$
Finally, 
by Markov's inequality,
the probability in the second term is 
no more than $k/n$, while the binary 
entropy is always bounded above by $1$.
Combining these three estimates yields,
\begin{equation*} 
I(X_i;X_{i+1}^k|A_\ell) 
\leq  \frac{2\ell(k-i)}{n(n-(k-i))}\log{n} 
+ \frac{k}{n} + 2\frac{k}{n-k}\log{n}.
\end{equation*}
The result follows. 
\qed

We are now ready to prove the theorem.
By the bound in the lemma,
$$\sum_{i=1}^{k-1}
I(X_i;X_{i+1}^k|A_\ell) \leq \frac{5k^2\log n}{n-k}.$$
Also, by definition of the mutual information,
using the obvious notation $H(X|A)$ for the 
entropy of the conditional p.m.f.\ of $X$ given $A$,
\begin{align*}
\sum_{i=1}^{k-1}I(X_i;X_{i+1}^k|A_\ell) 
&= 
	\sum_{i=1}^{k-1}\left[H(X_i|A_\ell) 
	+ H(X_{i+1}^k|A_\ell) 
	- H(X_i^k|A_\ell)\right] \\
&
	= \left[\sum_{i=1}^{k}H(X_i|A_\ell)\right]
	- H(X_1^k|A_\ell) \\
&= 
	D\big(Q_{X_1^k|A_\ell}\big\|
	Q_{X_1|A_\ell}\times\cdots\times Q_{X_k|A_\ell}\big),
\end{align*}
where we write $Q_{X_i^j|A_\ell}$ for the conditional p.m.f.\
of $X_i^j$ given $A_\ell$.
Since $Q_{X_i|A_\ell}=P_{\ell/n}$,
we have,
$$D\big(Q_{X_1^k|A_\ell}\big\|P_{\ell/n}^k)\leq\frac{5k^2\log n}{n-k}.$$
Finally, writing $\mu$ for the distribution of $\ell/n=(1/n)\sum_{i=1}^nX_i$
on $\{0,1/n,2/n\ldots,1\}$, averaging both sides with respect to $\ell$,
and using the joint convexity of relative entropy,
yields the claimed result.
\qed

\noindent
{\bf Remarks. }
The mixing measure $\mu=\mu_n$ in the theorem is completely characterised
in the proof as the distribution of 
$(1/n)\sum_{i=1}^nX_i$, and it is the same for all $k$.
Moreover, if $\{X_n\;;\;n\geq 1\}$ is an infinite 
exchangeable sequence then it is also stationary,
so by the ergodic theorem
$(1/n)\sum_{i=1}^nX_i$ converges a.s.\ to some~$X$,
and the $\mu_n$ converge weakly to the law, say $\mu$, of $X$.
For fixed $k$, 
since $P^k_p$ is a bounded and continuous
function of $p \in [0,1],$ 
we have for any $x_1^k \in \{0,1\}^k$, 
$$
M_{k,\mu_n}(x_1^k)=\int{P^k_p(x_1^k) d\mu_n(p)} 
\to M_{k,\mu}(x_1^k)=\int{P^k_p(x_1^k) d\mu(p)},
$$
and by our theorem,
$\|Q_k-M_{n,k}\|=O(\sqrt{(\log n)/n}).$
Therefore, we can conclude that, 
$$
Q_k = \int{P^k_p d\mu(p)},
$$
for each $k\geq 1$,
and thus recover de Finetti's classical representation theorem.

Finally we note that the argument used in the proof
of the lemma as well as the proof of our theorem can 
easily be extended to provide corresponding results
for exchangeable vectors taking values in any finite
set. But as the the constants involved become quite 
cumbersome and our main motivation is to illustrate
the connection with information-theoretic ideas
(rather to obtain the most general possible results),
we have chosen to restrict attention to the
binary case.

\medskip

\noindent
{\bf Acknowledgments. } We thank
Sergio Verd\'{u}
and an anonymous referee for useful 
suggestions regarding the presentation
of the results in this paper. 


{\small
\bibliographystyle{plain}

\def\cprime{$'$}

}

\end{document}